\documentstyle[aps,preprint,epsfig,eqsecnum]{revtex}
\input epsf

\begin{document}
\title{Conformations of Closed DNA}
\author{Boris Fain\cite{current} and Joseph Rudnick}
\address{Department of Physics,
UCLA \\ Los Angeles CA 90095-1547}
\maketitle
\begin{abstract}
We examine the conformations of a model for a short segment of closed
DNA. The molecule is represented as a cylindrically symmetric elastic
rod with a constraint corresponding to a specification of the linking
number.  We obtain analytic expressions leading to the spatial
configuration of a family of solutions representing distortions that
interpolate between the circular form of DNA and a figure-eight form
that represents the onset of interwinding.  We are also able to
generate knotted loops. We suggest ways to use our
approach to produce other configurations relevant to studies of DNA
structure.  The stability of the distorted configurations is assessed,
along with the effects of fluctuations on the free energy of the
various configurations.
\end{abstract}
\pacs{87.15.By, 62.20.Dc}

\section{introduction}

If a segment of untwisted rod is forced to close smoothly on itself,
it will take on a circular shape in order to minimize the elastic
energy.  Thermal fluctuations will exert a significant effect on this
configuration only if the circumference of the circle is greater than
the temperature-dependent persistence length of this segment.  When
the looped rod is also forced to undergo a twist, the circular shape
may prove to be unstable.  The rod will then either distort into a
non-planar, non-intersecting form, or it may wind about itself, in an
interwound, or ``plectonemic,'' configuration
\cite{plec1,plec2,plec3,plec4,plec5,plec6,plec7}.

When torsional and flexing stresses are not too severe,
\cite{bensimon} the DNA molecule can be modeled as such a rod.  A
short enough segment of looped DNA will thus take on shapes determined
by the minization of elastic energy.  Planar shapes, characteristic of
``relaxed'' DNA have been observed, as have the plectonemic forms when
the molecule is over-or underwound.  The distortion, or supercoiling,
of DNA under the influence of torsional stresses is widely believed to
have significant implications with respect to the action of this
molecule in biological settings \cite{sinden}.  That this is so may be
inferred from the existence of enzymes known as topoisomerases, which
act to facilitate the alteration of the torsional characteristics of
DNA configurations \cite{sinden,drlica,wang,gellert}.

The elastic model of DNA has been the subject of intense research in
the past 30 years.  Theoretical appraches to this problem include
Lagrangian mechanics\cite{benham,lebret,tsu,tanak,manning}, b-splines
and (numerical) molecular dynamics \cite{olsen}, and statistical
mechanics\cite{siggia}.  Only recently has it been possible to
determine the equilibrium configuration of a closed segment of DNA, as
modeled by distorted rod\cite{julicher}.  A comprehensive discussion
of the properties of these equilibrium configurations has not yet been
published.

In this paper we contribute to the knowledge of this configuration by
describing a method by which one can obtain the shape of a closed
section of an elastic rod, along with various key properties of this
system, including its twist, writhe, linking number and energy.  We
will use this formalism to generate both unknotted and knotted forms
of closed DNA. We will also outline how our formalism can be exploited
to produce other shapes of interest to DNA researchers.

The configurations that will be discussed include those that arise as
a result of the the supercoiling instability for a closed loop.  We
will also examine the configurations that represent the evolution of
higher energy deformation of a circular loop.  In addition, we will
briefly discuss configurations that are associated with knotted loops.
Such knots include, but are not limited to, the trefoil.  All the more
elaborate configurations are unstable with respect to fluctuations
about the solutions that extremize the energy of the loop.  However,
it is possible to envision mechanisms, analogous to those leading to
the nuclesome structure \cite{sinden,mcb}, that stabilize these
configurations and, therefore, cause them to be relevant to biological
systems.

As noted about, we will discuss the mechanical stability of the
configurations to be studied, and will present results with regard
to the contribution to the free energy of thermal fluctuations about
the minimizing shape.  These results will be relevant as long as
thermal fluctuations do not give rise to a significantly alteration of
the shape that the rod takes, and, as noted above, if the circumference
of the rod is not large compared to the rod's persistence length.  The
precise criterion for this will be examined later on in this paper.

The most important advance in the work presented here is the algorithm
for the generation of various closed structures and the analysis of
the stability of the configurations, particularly the one associated
with the principal supercoiling instability.  On the basis of the
stability analysis, it is clear that only some of the configurations
that one can generate with the use of the methods described in this
paper are, in and of themselves, mechanically stable.  However, the
existence of devices that support non-interwound supercoiled
configurations in the case of open strands of DNA \cite{sinden,mcb}
allow one to speculate as to the possible relevance of those
configurations to DNA in biological settings.  Furthermore, one can
envision utilizing these methods in conjunction with other
calculational tools to study the mechanical properties and packing
configurations of longer strands of DNA.

The paper is organized as follows.  First, the mechanical model is
described and important parameters are defined.  Of especial
importance is an identification of the parameters that will be
adjusted in order to obtain solutions of the energy extremum
equations.  Key quantities, including topological invariants, will be
defined in terms of those parameters.  The stability of the circular
configuration of a closed rod under torsional stress is then assessed.
The principal instability is identified, and it is shown how to obtain
the family of closed curves that represent the evolution of this
instability.  This family interpolates between the circle and a figure
eight, that latter of which lies at the onset of interwinding.  The
stability of the members of this class is then assessed.  It is found
that they are, indeed, stable against small fluctuations.  Then, it is
shown that the method by which the outcome of the supercoiling
instability is investigated can be utilized to generate deformed
configurations, among which are configurations in which the closed
loop is also knotted.  Finally, there is a brief discussion of the way
in which the contributions of fluctuations about the classical
solution configurations.  Details and background are relegated to
appendices.

\section{Elastic Model and Expressions of Interest}

In previous work we have outlined some developments in the elastic
model of DNA\cite{fain}.  The molecule is represented as a slender
cylindrical elastic rod.  At each point $s$ the rod is characterized
by relating the local coordinate frame ${\cal F}$ to the frame ${\cal
F}_0$ rigidly embedded in the curve in its relaxed configuration.  The
relationship between the stressed and unstressed local frames is
specified by Euler angles $\theta(s), \phi(s), \psi(s)$ needed to
rotate ${\cal F}_0$ into $\cal F$.

We begin by summarizing some of the results needed for this paper.  We
denote the elastic constants of bending and torsional stiffness by $A$
and $C$, respectively.  The length of the rod is denoted by $L$.  The
elastic energy is given by
\begin{equation}
E_{el}=\int_{0}^{L}ds\;\left(
\frac{A}{2}\left(\dot{\phi}^2\sin^2\theta+\dot{\theta}^2\right) +
\frac{C}{2}\left(\dot{\phi}\cos\theta+\dot{\psi}\right)^2\right)
\label{c_energy_eqn}
\end{equation}
The ``dot'' notation for derivatives is used interchangeably with the
explicit derivative with respect to arc-length, $d/ds$.

The twist is given by
\begin{equation}
{\cal T}w=\frac{1}{2\pi}\int_{0}^{L}ds\;
\left(\dot{\phi}\cos\theta+\dot{\psi}\right) \label{tw_eqn}
\end{equation}
A relevant result from our previous work is that
${\cal  L}k$, the linking number, can be written as
\begin{equation}
{\cal L}k= {\cal I} - 1 +\frac{1}{2\pi}\int_{0}^{L}\left(\dot{\phi}
+\dot{\psi}\right)\;ds  \label{lk_eqn}
\end{equation}
Where we have used Fuller's theorem\cite{fuller} to obtain the writhe,
and White's theorem\cite{white} for the total link. The quantity
${\cal I}$ is an integer that encodes the knotting of the loop. It is
equal to the number of signed crossings of the loop, assuming that it
has been flattened out as much as possible.

\section{Candidates: The Elastic Extrema}

\subsection{Stability and Excitations of the Circular Loop}

As a precursor to the discussion of the possible deformations of a
closed loop, we review the family of excitations of a closed, circular
loop, and the conditions under which this configurations may become
unstable to one or more of these excitations.  Detailed discussion of
these issues has been relegated to Appendices \ref{sec:genstab} and
\ref{sec:circfluc}.  The Appendices also contain derivations of key
formulas presented below.

The excitation of a circular loop is describable in terms of the
deviation of the Euler angle $\theta$ from the constant value of $\pi/2$
that it has in the case of a circle. If we write $\theta(s) = \pi/2 +
\gamma(s)$, then small excitations about a circular shape will be of
the form
\begin{equation}
\gamma(s) \propto \cos \left( \frac{ 2 \pi n}{L} + \delta \right)
	\label{flucform1}
\end{equation}
where $n$ is an integer (possibly 0) and $\delta$ is an arbitrary
phase factor.  The energy of such a sinusoidal excitation proportional
to
\begin{equation}
E_{n} = \frac{A}{2}\left(\frac{2
\pi}{L}\right)^{2}\left\{n^{2}-1-\left(\frac{C}{A}\right)^{2}{\cal
L}k^{2} \right\}
	\label{enform1}
\end{equation}

The requirement that the fluctuations leave the loop closed and, also,
that they not change the linking number of the loop rule out $n=0$ and
$n=1$ as possibilities.  The allowed deformation with the smallest
energy has $n=2$. Substituting this value into Eq. (\ref{enform1}), we
find for the energy cost of this excitation
\begin{equation}
E_{2} \propto \frac{A}{2}\left(\frac{2 \pi}{L}\right)^{2}\left\{ 3 -
\left(\frac{C}{A}\right)^{2}{\cal L}k^{2} \right\}
	\label{enform2}
\end{equation}
It is clear that this deformation of the circle can give rise to a
lowering of the energy. This state of affairs holds if
\begin{equation}
{\cal L}k  > \frac{A}{C}\sqrt{3}
	\label{supcoil1}
\end{equation}
In fact, the condition in Eq. (\ref{supcoil1}) is just the
requirement for the supercoiling instability of the closed loop.
Deformations associated with higher values of $n$ lower the energy of
the loop when the condition
\begin{equation}
{\cal L}k > \frac{A}{C}\sqrt{n^{2}-1}
	\label{supcoil2}
\end{equation}
holds.

It is clear that the instability with the earliest onset, and the one
that will prove dynamically ``strongest'' is the one with threshold as
given by (\ref{supcoil1}).

\subsection{A Family of Curves}

We start by looking for a family of closed solutions to the
Euler-Lagrange equations that represent the evolution of the
deformation associated with the supercoiling instability, \emph{i.e.}
$n=2$ in Eqs (\ref{flucform1}) and (\ref{enform1}) above.  This family
of curves can be indexed by the writhe, ${\cal W}r$, of a member,
which ranges from ${\cal W}r=0$ for a simple circle to ${\cal W}r=1$
of the other limiting member of the family, a ``figure eight.''  The
actual constraint imposed on the closed curves that are of relevance
to this discussion is that the link is fixed at a predetermined value.
We shall therefore examine the conditions under which the family of
writhing curves can satisfy the imposition of a linking number

\subsection{Euler-Lagrange Minimization}
Our goal is to extremize elastic energy and keep the curve closed.
The functional to consider for closed configurations of DNA is
\begin{equation}
{\cal H} = \int_{0}^{L}dE_{el} - F{\bf t}{\bf{\hat z}}\:ds
 \label{c_hamil_eqn}
\end{equation}
The first term of (\ref{c_hamil_eqn}) is clear---the elastic energy
must be extremized.  The second term enforces the constraint that the
curve minimizing the elastic energy also closes upon itself.  In
general, one expects to introduce Lagrange multipliers controlling the
extent of the curve in all directions.  However, one can concentrate
on the net displacement in only the $z$ direction, because it is
always possible to transform to a a reference frame in which the curve
is closed in the $XY$ plane.  Written explicitly as a function of
Euler angles (\ref{c_hamil_eqn}) becomes
\begin{eqnarray}
{\cal H} = \int_{0}^{L}\;ds \hspace{0.2in}
&&\frac{A}{2}\left(\dot{\phi}^2\sin^2\theta+\dot{\theta}^2\right)+
\nonumber \\
 +&&\frac{C}{2}\left(\dot{\phi}\cos\theta+\dot{\psi}\right)^2
 -F\cos\theta  \label{c_hamiltonian}
\end{eqnarray}
The extrema are found by applying Euler-Lagrange equations
to (\ref{c_hamiltonian}). Denoting the conserved quantities as
$J_{\phi} \equiv \partial{\cal H}/\partial \dot \phi$
and $J_{\psi} \equiv \partial{\cal H}/\partial \dot \psi$
we obtain
\begin{eqnarray}
\dot\phi&=&\frac{J_\phi-J_\psi\cos\theta}{A\sin\theta}
\label{c_phi_eqn}\\
\dot\psi&=&\frac{J_\psi}{C}-\dot\phi\cos\theta
\label{c_psi_eqn}
\end{eqnarray}
The equation for $\theta$ is a quadrature obtained by
integrating $\partial{\cal H}/\partial \theta
=d/ds \partial{\cal H}/\partial \dot \theta$
with $E_0$ as the constant of integration.
Defining $u\equiv\cos\theta$ we see that the behavior of
solutions is governed by a cubic polynomial in $u$:
\begin{eqnarray}
{\dot u}^2&=&\frac{2\left(1-u^2\right)}{A}
\left(E_0-Fu\right)-
\frac{1}{A^2}\left(J_\phi^2+J_\psi^2-2 J_\phi J_\psi u \right)
\nonumber \\
&\equiv&\frac{2F}{A}\left(u-a\right)\left(u-b\right)\left(u-c\right)
\label{theta_eqn}
\end{eqnarray}
where we order the roots $c\leq u(\equiv\cos\theta(s))\leq b\leq a$.
Eq.  (\ref{theta_eqn}) requires that $\cos\theta(s)$ oscillates
between $c$ and $b$.  For the distortions studied here, there are two
complete ocillations as the the closed curve is traversed once.  All
the relevant quantities, including the shape of the curve, can be
obtained using (\ref{theta_eqn}).  For example:
\begin{eqnarray}
\phi(s)&=&\int_{u(s=0)}^{u(s)}\frac{d\phi}{ds}\frac{ds}{du}du
\;\;\;\mbox{ with } \nonumber\\
\frac{ds}{du}&=&\sqrt{\frac{A}{2F}}
\frac{1}{\sqrt{\left(u-a\right)\left(u-b\right)\left(u-c\right)}}
\label{c_quadrature_eqn}
\end{eqnarray}
The key to the solution of the equation for the distorted closed loop
is the determination of the parameters $F$, $a$, $b$ and $c$.

\subsection{Constraints}
\begin{figure}
\centerline{\epsfig{file=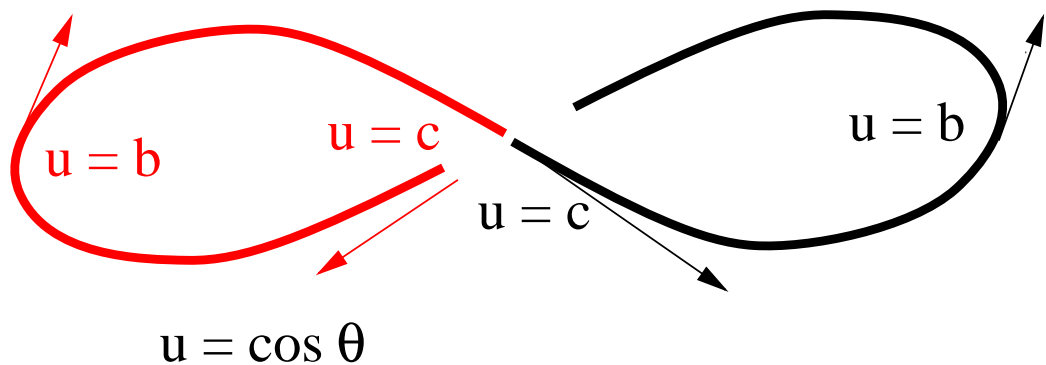,height=3in}}
\caption{
The tangent oscillates $c\rightarrow b\rightarrow c\rightarrow b
\rightarrow c$.
 The curve is composed
of two symmetric parts.}
\label{segment_fig}
\end{figure}

To complete the analysis we must determine the parameters generated by
the minimization procedure.  One choice of parameters is the set of
invariants $J_\phi$ and $J_\psi$, the constant $E_0$, and the Lagrange
multiplier $F$.  However, as indicated above, a better practical
choice is the Lagrange multiplier $F$ and the parameters $a$, $b$ and
$c$. To determine the parameters we impose constraints on
the curve.  Figure \ref{segment_fig} is a good visual guide to the
geometric meaning of the constraints. They are
\begin{enumerate}
\item[A] The loop closes on itself in the $z$-direction. Because the
curve consists of four segments, in each of which the variable $u(s)$
goes from $b$ to $c$, or from $c$ to $b$, we have
\begin{equation}
\int_{0}^{L/4}u(s)ds = 0
	\label{zclosure}
\end{equation}

\item[B] The loop goes through one quarter of a turn in each segment.
This reduces to
\begin{equation}
\int_{0}^{L/4}\frac{d \phi(s)}{ds}ds = \frac{\pi}{2}
	\label{phiequation}
\end{equation}

\item[C] The linking number takes on a predetermined value for the loop.
This requirement leads to the following mathematical constraint on the
solution to the Euler Lagrange equations.
\begin{equation}
\int_{0}^{L} \frac{d \psi(s)}{ds}ds = {\cal L}k
	\label{linkconstraint}
\end{equation}

\item[D] Finally, the length of the loop takes on a predetermined
value. Mathematically,
\begin{equation}
\int_{0}^{L}ds = L
	\label{lengthconstraint1}
\end{equation}
The above condition is not as tautological as it seems. The actual
parameterization of the curve will be in terms of the dependence of
quantities on the cosine of the Euler angle $\theta(s)$.
\end{enumerate}

\subsection{Reparameterization}

At this point we reparameterize the problem in terms of $F,a,b,c$
instead of $F, E_0, J_\phi, J_\psi$.  Parametrizing the problem by the
roots of the polynomial is extremely advantageous: it makes the
analytic manipulations more transparent; it also streamlines the
computational tasks.  The two sets of parameters are related in the
following manner:
\begin{eqnarray}
E_0&=&F\left(a+b+c\right))  \label{Edef}\\
J_\phi&=&\sqrt{\frac{AF}{2}}\left( p_1 \pm p_2\right);\;\;\;
J_\psi=\sqrt{\frac{AF}{2}}\left( p_1 \mp p_2\right) \label{Jdefs} \\
&\mbox{with}&\;p_{1\choose 2}\equiv\left[(c \pm 1)(b \pm 1)(a \pm
1)\right]^{1/2}
\nonumber
\end{eqnarray}
The choice of branch ($\pm$) is imposed by the family of
configurations sought.  For circular DNA without intrinsic curvature,
$J_\phi$ takes a $-$, and $J_\psi$ correspondingly a $+$.\footnote{the
choice of a particular branch is a non-trivial procedure} Let us make
some definitions that render the notation more transparent:
\begin{eqnarray}
&&\Delta b\equiv \left(b-c\right)\;\; ; \;\; \Delta a\equiv
\left(a-c\right) \label{delta_eqn}\\
&&q\equiv\sqrt{\frac{\Delta b}{\Delta a}} \label{q_eqn}
\end{eqnarray}
Employing (\ref{c_quadrature_eqn}) we rewrite the constraint
equations, (\ref{zclosure}), (\ref{phiequation}), and
(\ref{lengthconstraint1}), in terms of the new parameters:
\begin{eqnarray}
&&aK\left(q\right)=\Delta aE\left( q \right) \label{c_a_constraint}\\
&&\pi/2\;\;=\;\;\frac{1}{\Delta a}\left[
\left|\frac{(a+1)(b+1)}{(c+1)}\right|^{1/2} \Pi\left(\frac{\Delta
b}{-1-c},q \right)\right.- \nonumber \\
&&\hspace{2.15cm}-\left.\left|\frac{(a-1)(b-1)}{(c-1)}\right|^{1/2}
\Pi\left(\frac{\Delta b}{1-c},q \right)\right]
\label{c_phi_constraint}\\
&&F=\frac{32A}{\Delta a L^2}K\left(q\right)\;\;\;\;
\label{c_f_constraint}
\end{eqnarray}
Where $K,E\mbox{ and }\Pi$ are complete elliptic integrals of the
first, second and third kind, respectively. The issue of the
constraint on linking number will be left for later discussion.

\subsection{Solving for $a$, $b$, $c$ and $F$}

We find that the optimal procedure for calculation of parameters
appropriate to a solution is to utilize (\ref{c_f_constraint}) to
eliminate $F$, then (\ref{c_a_constraint}) and
(\ref{c_phi_constraint}) eliminate $\Delta a$ and $c$.  Because the
writhe, ${\cal W}r$, is a monotonically increasing function of $\Delta
b$, we make use of this property to classify distinguish between the
members of a family of curves.  Once the constants $F,a,b,c$ are
determined, the desired solutions and all the relevant quantities are
computed via elliptic integrals.  For example, the explicit expression
for $\theta(s)$ in the first quarter of oscillation is (we have
inverted (\ref{c_quadrature_eqn})):
\begin{equation}
\cos\theta=\Delta b \mbox{ sn}^2\left(\sqrt{\frac{F\Delta c}{2A}}s,
\sqrt{\frac{\Delta b}{\Delta a}}\right)-c \label{c_explicit_theta_eqn}
\end{equation}
$\phi \mbox{ and }\psi$ are obtained similarly from (\ref{c_phi_eqn})
and (\ref{c_psi_eqn}).

\begin{figure}
\centerline{\epsfig{file=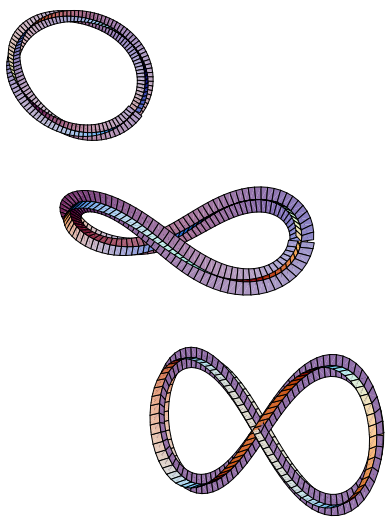,height=3in}}
\caption{The family of curves ranges from the circle in the $XY$ plane
to the figure eight in the $YZ$ plane.  The perspective is slightly
asymmetric to aid visualisation.  The ``fins'' on these, and all other
pictured curves, trace out the imbedded $x$ and $y$ axes, and thus
depict the twisting of the rod.}
\label{circle_family_fig}
\end{figure}

\noindent Figure \ref{circle_family_fig} displays the family of curves
computed in the
manner discussed above. Since the constraint equations involve
elliptic integrals, finding a solution on a computer is virtually
instanteneous; analytically and computationally elliptic
integrals are equivalent to, say, $\arcsin$.

\subsection{Bounding Members: Circle and Figure Eight}
Let us check whether the initial member of our family, a
circle with ${\cal W}r=0$ joins smoothly with the previously
known stable family of twisted circles\cite{benham_stab}.
The circle corresponds to ${\cal W}r=0$.
A circle in the $XY$ plane the curve must have
$b_0=c_0=0$. (\ref{c_phi_constraint}) now states that

\begin{eqnarray}
&&\frac{1}{\sqrt a_0}\left(\sqrt{a_0+1}-\sqrt{a_0-1}\right)=1\\
&&\mbox{which in turn gives}\nonumber\\
&&{J_\psi}_0=\frac{2\pi A}{L}\frac{1}{\sqrt a_0}
\left(\sqrt{a_0+1}+\sqrt{a_0-1}\right)=\frac{2 \pi A}{L}\sqrt{3}
\label{p_circle_equation}
\end{eqnarray}
Combining (\ref{tw_eqn}) and (\ref{c_psi_eqn}) to obtain the twist, ${\cal
T}w$, of the circle (\ref{p_circle_equation}) gives
\begin{equation}
{\cal T}w_0=\frac{L}{2\pi}{J_\psi}_0=\sqrt{3}\frac{A}{C}
\end{equation}

It is also of interest to compute the ${\cal T}w$ of the figure
eight, which, like the circle, can be performed virtually by inspection.
The figure lies in the $YZ$ plane, which forces $\phi$ to behave as
follows: (refer to Figure \ref{segment_fig} for visualization)
\begin{equation}
{\dot{\phi}}_8=\pi\left(\delta\left(0\right)+\delta\left(\frac{L}{2}
\right)\right)
\label{c_phi_eight_eqn}
\end{equation}
Combining (\ref{c_phi_eight_eqn}) and (\ref{c_phi_eqn}) forces
${J_\phi}_8={J_\psi}_8=0$ which immediately sets ${\cal T}w_8=0$.  The
value of $\Delta b_8$ is easily determined from the fact that $c_8=-1$
(this can be seen from the curve itself) necessitating $a=1$.  Then
(\ref{c_phi_constraint}) yields $\Delta b_8=1.6522\ldots$.

\section{Linking Number and the Plectonemic Transition}
We have found a family of writhing solutions that are the extrema of
elastic energy.  The writhe of the curves covers the range $0\leq{\cal
W}r\leq1$.  \footnote{That the figure eight has ${\cal W}r=1$ just
before crossing can be seen from the shape of the curve.}
However, the physical constraint imposed on the molecule is ${\cal
L}k$, the linking number.  The explicit expressions for ${\cal T}w$,
${\cal L}k$ and ${\cal W}r$ are easily obtained:
\begin{eqnarray}
{\cal T}w&=&\frac{2\pi}{\sqrt{\Delta a}}K\left(q\right)\left(p_1+p_2\right)
\frac{A}{C}\\
{\cal W}r&=&\frac{2\pi}{\sqrt{\Delta a}}\left[-K\left(q\right)\left(p_1+p_2
\right)\right.\nonumber\\
&+&\left.\frac{p_1}{1+c}\Pi\left(\frac{\Delta b}{-1-c},q \right)+
\frac{p_2}{1-c}\Pi\left(\frac{\Delta b}{1-c},q \right) \right]
\end{eqnarray}
and, using White's theorem\cite{white}
\begin{equation}
{\cal  L}k={\cal T}w+{\cal W}r
\end{equation}

\begin{figure}
\centerline{\epsfig{file=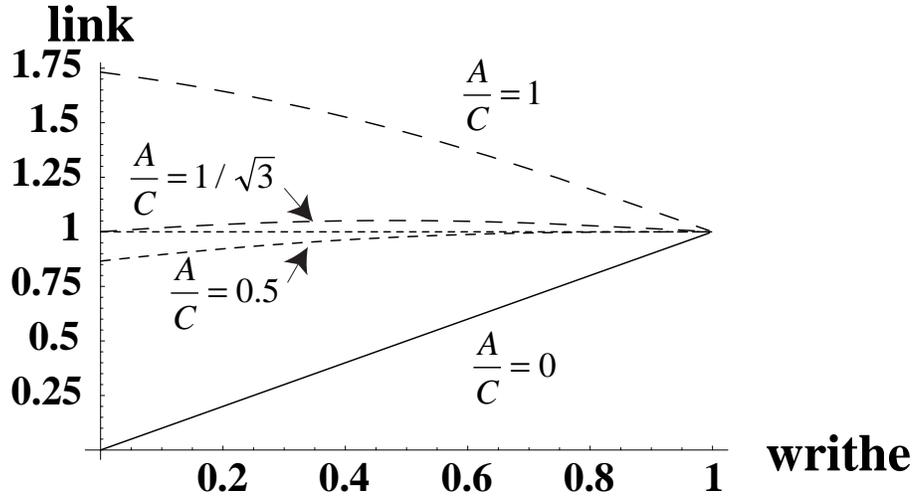,height=3in}}
\caption{
The three types of behaviour of ${\cal  L}k$ for our writhing
family of curves.
The ratio $A/C$ controls the plectonemic transition.}
\label{link_family_fig}
\end{figure}

With reference to Figure \ref{link_family_fig}, we can now describe
what happens to a deformed loop of DNA as the linking number is
increased.  The loop remains a circle until $\Delta {\cal
L}k=\sqrt{3}A/C$.  After that there are three possibilities.  If
$A/C\le 0.5$ then the writhing family supports a steady increase to
the linking number limit ${\cal L}k=1$ of the figure eight and the
molecule folds continuously until self-crossing occurs.  Further
increase of ${\cal L}k$ presumably results in a plectonemic
configuration.  If $0.5 \le A/C\le 1$ the one can expect the curve to
distort continuously until the writhe achieves some value intermediate
between zero and one.  There is, at this threshold value of the
writhe, a transition, almost certainly to an interwound form.  When
$A/C > 1$, none of the members of the writhing family support the
neccesary linking number, and as soon as $\Delta {\cal L}k$ exceeds
the supercoiling threshold value $\sqrt{3}A/C$ the twisted circle
snaps into a plectoneme.  An interesting fact is that this behavior is
{\em independent} of the length of the molecule.  That this ought to
be so follows from the absence of an absolute length scale in the
problem of the deformed loop.

Greater insight into the behavior of the deforming loop is gained if
one also investigates the way in which the energy depends on the
various properties of the loop. The energy can be
expressed in terms of the previously introduced quantities. A
straightforward calculation yields for the energy of the loop
\begin{equation}
E = F \left\{ \left(a+b+c \right) +
\frac{1}{4}\left(p_{1}+p_{2}\right)^{2} \left[\frac{A}{C}-1 \right]
\right\} \int ds
	\label{energy1}
\end{equation}
Where the final integral is over arclength. Expressing the
circumference of the loop in terms of $F$, $a$, $b$ and $c$, we
obtain the following result for the energy in terms of the total
circumference of the loop, $l$, the bending modulus, $A$, and the
parmeters $a$, $b$, and $c$:
\begin{equation}
E = \frac{2A}{l}
\left\{ \left(a+b+c \right) +
\frac{1}{4}\left(p_{1}+p_{2}\right)^{2} \left[\frac{A}{C}-1 \right]
\right\}
\left(\int_{c}^{b}\frac{du}{\sqrt{(a-u)(b-u)(u-c)}}\right)^{2}
	\label{energy2}
\end{equation}

\begin{figure}
\centerline{\epsfig{file=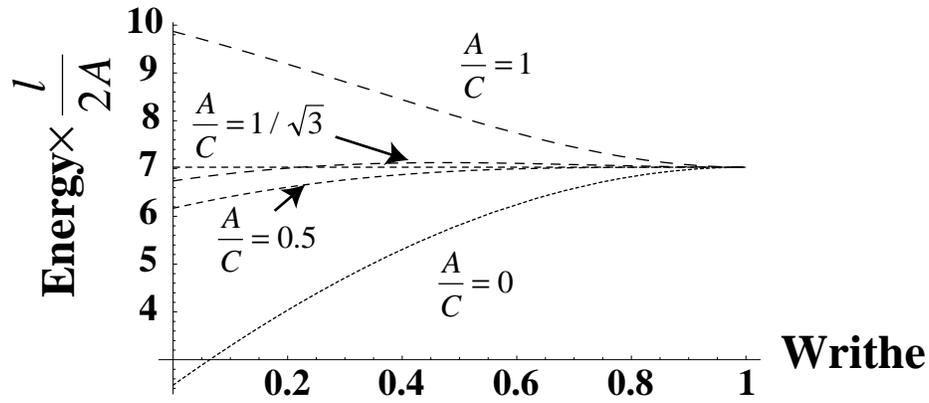,height=3in}}
\caption{The energy as a function of writhe for in the cases plotted
in Figure \ref{link_family_fig}. The energy is expressed in units of
the combination $2A/l$, where $l$ is the circumference of the writhing
loop, and $A$ is the bending modulus. Also shown in this plot is the
limiting value of the energy, at the figure eight configuration,
which is displayed as a horizontal line.}
\label{fig:energyplot}
\end{figure}

The energy of the loops as a function of writhe for various values of
the modulus ratio $A/C$ is plotted in Figure \ref{fig:energyplot}.  A
few facts about the energies of the deformed configuration can be
established numerically (and we do not doubt that analytic
demonstrations can also be constructed).  First, the energy is a
monotonically increasing function of linking number, if not of writhe.
This is not immediately evident in Figure \ref{fig:energyplot},
although it is indicated, in that the energy increases or decreases
monotonically with writhe when the link does so.  Furthermore, when
there is a maximum in link as a function of writhe, there is also an
energy maximum, and, as indicated in Figures \ref{link_family_fig} and
\ref{fig:energyplot}, the maxima occur at the same value of the
writhe.

In addition, when there is a maximum in the linking number as the
writhe increases from zero to one, so that there are two possible
writhes associated with the same linking number in for a range of the
latter quantity, the configuration with the lower writhe always has
the lower energy. This can be established by a separate set of
calculations.  This  implies that when, for instance, $0.5 < A/C <
1$, the loop, in distorting from its planar form, will not be
susceptible to a discontinuous distortion to a more highly writhed,
but not interwound, configuration before it ``snaps'' to a
plectoneme. In the absence of thermal fluctuations or external
perturbations, that transition will to occur when the link has
reached its maximum value as a function of writhe.

For a more detailed discussion of the energetics of the deforming
loop, along with a calcuation of the energy of the plectoneme, the
reader is referred to \cite{julicher}

\section{Stability of the deformed loop}
\label{sec:loopstab}

Given the existence of the solutions for the deformed loop
corresponding to the evolution of the supercoiling instability, it is
desirable to determine whether these solutions are, themselves, stable
against small fluctuations.  There is no {\em a priori} guarantee that
this is so.  We will find, in fact, that the stability of the closed
configuration is enforced by the relatively large number of
constraints that must be satisfied by fluctuations about the
extremizing solutions to the Euler-Lagrange equations.  A number of
technical details associated with the determination of the stability
of a configuration are to be found in various appendices.

\subsection{Stability in the Absence of Constraints: the Translation
Mode}

\begin{figure}
\centerline{\epsfig{file=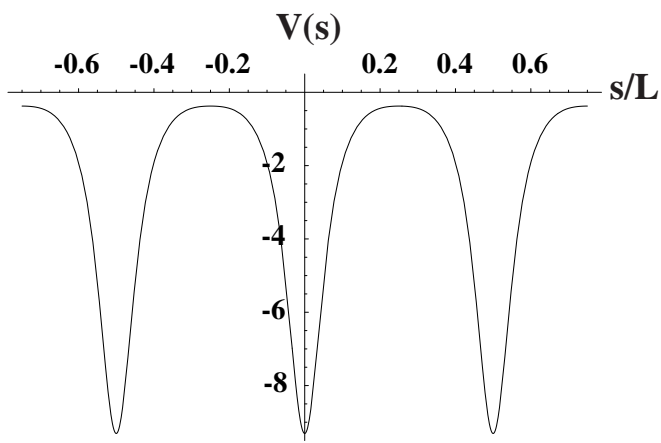,height=3in}}
\caption{Typical potential, as given by Eq.  (\ref{potential1}), in
the Schr\"{o}dinger-like equation, (\ref{schrod1}), obeyed by
fluctuations about the deformed state.  The quantity $L$ is the total
arclength of the loop.}
\label{fig:pot}
\end{figure}

The energy of a fluctuation about a solution to the Euler-Lagrange
equation is expressible in terms of solutions to the linear second
order differential equation (\ref{schrod1}), equivalent to the
Schr\"{o}dinger equation of a particle in the potential
(\ref{potential1}). In particular, any fluctuation, $\zeta(s)$ of the
Euler angle $\theta(s)$  about the form it takes in an
extremizing solution can be written in the form
\begin{equation}
\zeta(s) = \sum_{l}K_{l}\Psi_{l}(s)
	\label{zeta1}
\end{equation}
where $\Psi_{l}(s)$ is a solution to Eq. (\ref{schrod1}), with
eigenvalue $\lambda_{l}$. Assuming that the $\Psi_{l}$'s are
normalized, the energy of this fluctuation is given by
\begin{equation}
\sum_{l}K_{l}^{2}\lambda_{l}
	\label{enexpansion}
\end{equation}
The extremizing solution will be stable as long as there are no
solutions to Eq.  (\ref{schrod1}) with negative eigenvalues.  Now,
consider the potential $V(s)$, displayed in Figure \ref{fig:pot},
associated with one particular member of the family of solutions that
we have been considering.

\begin{figure}
\centerline{\epsfig{file=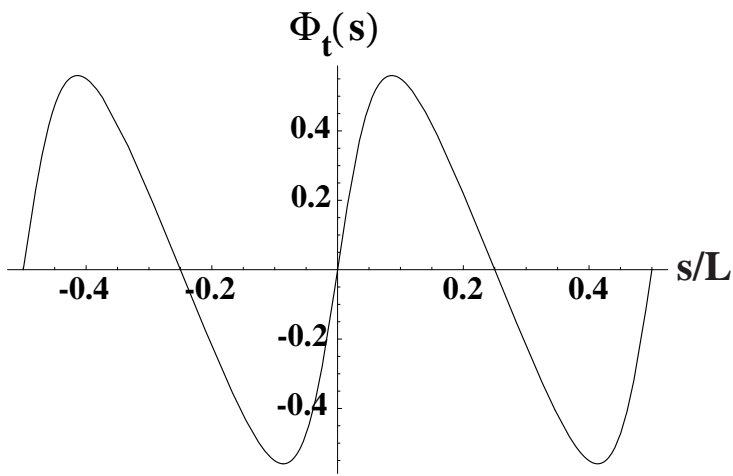,height=3in}}
\caption{The translational mode, $\Phi_{t}(s)$. The normalization of
the mode is such that its derivative is unity at $s=0$. The quantity
$L$ is the total arclength of the loop.}
\label{fig:trans}
\end{figure}

A striking property of this potential is that it is always negative.
From this one can infer that there are solutions of the
Schr\"{o}dinger-like equation with negative eigenvalues.  In fact, the
existence of such states is guaranteed by the existence of the
translational mode---the ``fluctuation'' equal to the derivative with
respect to arc length of the extremizing $\theta(s)$.  As the system
is invariant with respect to translations along the closed loop, the
infinitesimal transformation of the extremizing solution
$\theta_{cl}(s) \rightarrow \theta_{cl}(s + \delta s) = \theta_{cl}(s)
+ \delta s d \theta_{cl}(s)/ds$ has no effect on the energy of the
configuration.  The function $\Phi_{t}(s) \propto d \theta_{cl}(s)
/ds$ has the requisite periodicity, in that it is unchanged if $s$
goes to $s \pm n L$.  This implies the existence of a solution to the
Schr\"{o}dinger equation with zero eigenvalue.  This solution is
displayed in Figure \ref{fig:trans}.  In both Figure \ref{fig:pot} and
\ref{fig:trans}, the arclength parameter ranges from $-L/2$ to $L/2$.
This will prove to be useful later on, when we take advantage of the
reflection invariance of the portential $V(s)$.

The important thing to note is that the translational mode has
nodes. Note, furthermore, that this mode is odd on reflection about
$s=0$. On the basis of a simple node-counting argument, one can
readily establish that there will be another antisymmetric solution to
the equation (\ref{schrod1}), having fewer nodes in the interval
$-L/2<s<L/2$, and, hence, a lower eigenvalue than the translational mode.
This lower eigenvalue is necessarily negative. As it turns out, there
are two symmetric solutions to (\ref{schrod1}) having the requisite
periodicity and negative $\lambda$'s. The extremizing solution is,
thus, nominally unstable with respect to {\em three} kinds of
fluctuations.

\subsection{Effects of Constraints}

Certain constraints apply to any fluctuation in a closed loop.  In
fact, there are five such constraints, listed in Appendix
\ref{sec:genstab}.  These constraints must be incorporated into any
calculation of the stability of an extremizing solution.  The general
effect of such constraints on the stability calculation is outlined in
Appendix \ref{sec:genconst}.  The determination of the stability of a
deformed loop amounts to a search for zeros of a determinant obtained
by sandwiching a response kernel between five different functions,
each associated with one of the five constraints that must be
satisfied by any small variation of the solution to the Euler Lagrange
equations for the extremal loop configuration.  The five functions are
readily extracted from the integrals in Eqs.
(\ref{smooth1})--(\ref{zclosure2}).  With the use of symmetry
arguments, one can verify that the five-by-five matrix constructed out
of the response kernel, which is exhibited in Appendix \ref{sec:resp},
reduces to a four-by-for matrix and a one-by-one matrix.  The latter
matrix consists of the expectation value of the response kernel with
respect to the function associated with closure in the $y$ direction.
This function is contained in the integral in Eq.  (\ref{yclosure2}).

\begin{figure}
\centerline{\epsfig{file=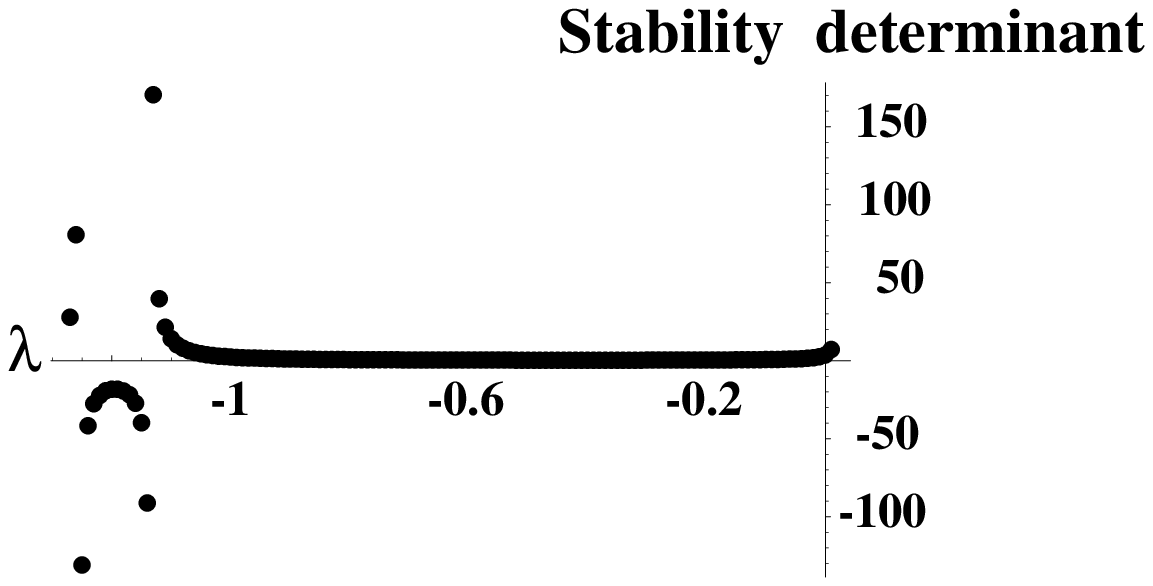, height=3in}}
\caption{The determinant of the four-by-four matrix obtained by
sandwiching the response kernel for the deformed loop between four of
the five functions associated with the constraints on fluctuations
about a solution of the Euler-Lagrange equations for a closed loop.
Here the parameter $a$, $b$ and $c$ are equal to 1.14903, 0.145102
and -0.154898, respectively. The poles in the determinant are at the
locations of the negative $\lambda$'s associated with unconstrained
fluctuations about the extremal solution.}
\label{fig:det1}
\end{figure}

\begin{figure}
\centerline{\epsfig{file=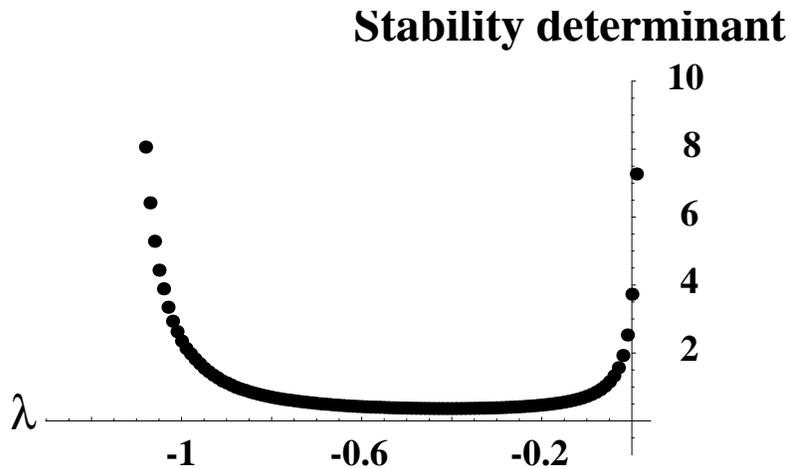, height=3in}}
\caption{A more detailed version of Fig \ref{fig:det1}, in which the
region between the pole at the less negative $\lambda$  and the origin
is displayed, to highlight the fact that the determinant does not go
through zero when $\lambda$ is negative.}
\label{fig:det1a}
\end{figure}

\begin{figure}
\centerline{\epsfig{file=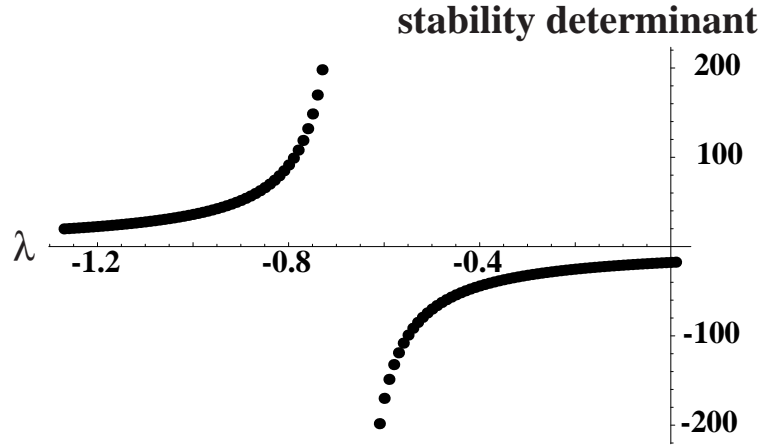,height=3in}}
\caption{The determinant of the one-by-one matrix obtained by
sandwiching the response kernel for the deformed loop between the
function associated with the constraint of closure in the
$y$-direction. The parameters $a$, $b$ and $c$ are equal to 1.14903, 0.145102
and -0.154898, respectively. The pole in the determinant is
at the location of the negative $\lambda$ associated with an
unconstrained fluctuation about the extremal function. }
\label{fig:det2}
\end{figure}

The matrices are straightforwardly constructed, although the
calculation is somewhat tedious.  We find that all the configurations
associated with the supercoiling instability that interpolate between
the circle and the figure eight are mechanically stable in that the
for no negative value of the parameter $\lambda$ does the determinant
pass through zero.  An example of the calculation of this determinant
for a particular deformation of the circular loop is displayed in
Figures \ref{fig:det1}, \ref{fig:det1a} and \ref{fig:det2}.

\section{Generalizations}

\subsection{Higher Order Deformations}

\begin{figure}
\centerline{\epsfig{file=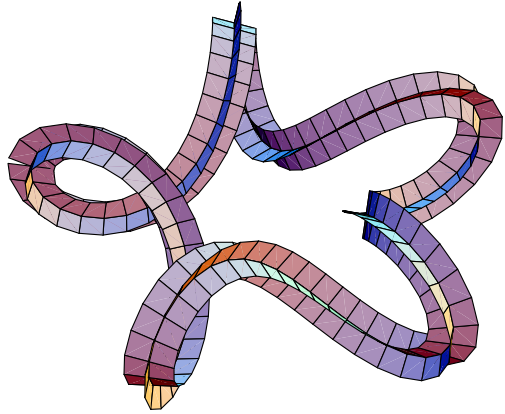,height=3in}}
\caption{The deformation with fivefold symmetry.  This represents one
of the most ``developed'' member of the family of fivefold deformed
loops.  The perspective is off-center to aid the visualization.  The
actual curve is fivefold symmetric. The ``fins'' on this curve trace
out the imbedded $x$ and $y$ axes, and thus depict the twisting of the
rod.}
\label{fig:fivefold}
\end{figure}

The solution that interpolates between the supercoiling instability
and the figure eight deformation bordering on the regime in which
the loop interwinds does not represent the only possibility for
deformation.  One can also generate deformed loops associated with the
evolution of higher energy excitations of the circular loop.  This is
accomplished by altering the requirement on the change in the Euler
angle $\phi$ over a complete ``period'' of the oscillation of the
variable $u(s)=\cos \theta(s)$, as it cycles between its limiting
values of $b$ and $c$.  If, instead of asking that $\phi$ advance by
$\pi/2$ in half a period, one requires a change of $\pi/n$, then it is
possible to generate a family of solutions to the Euler-Lagrange
equations associated with the excitations having $n$ periods around
the circumference of the circular loop.  Figure \ref{fig:fivefold}
displays such a distortion of the circular loop.  Here, $n=5$.  The
stability analysis of this this family of solutions is
straightforward.  Starting with the translational mode, one counts
nodes and determines the number of solutions to the eigenvalue
equation for fluctuations for which $\lambda$ must be negative.  There
are simply too many to allow for stabilization by the action of
constraints. In the absence of external stabilizing mechanisms, such
as histones or their equivalent, a loop will spontaneously distort
out of this configuration.

\subsection{Knotted Configurations}

\begin{figure}
\centerline{\epsfig{file=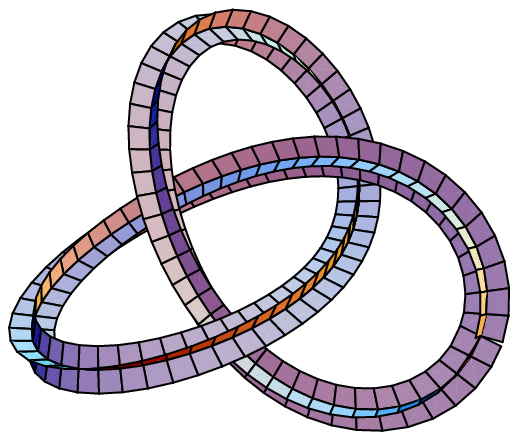,height=3in}}
\caption{A trefoil solution to the Euler-Lagrange equations.This is a
member of a family of solutions to the Euler Lagrange equations that
knots in this particular way.  It represents one of the more strongly
``writhed'' members of this particular class of solutions.The ``fins''
on this curve trace out the imbedded $x$ and $y$ axes, and thus depict
the twisting of the rod.}
\label{fig:trefoil}
\end{figure}

\begin{figure}
\centerline{\epsfig{file=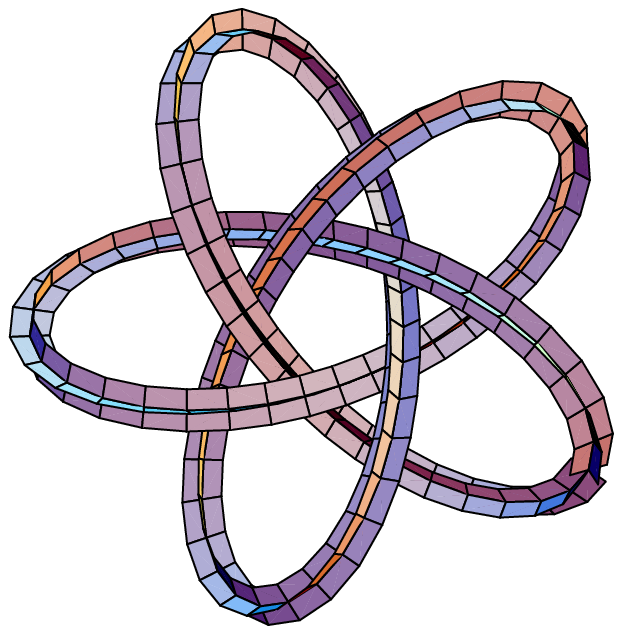,height=3in}}
\caption{A fivefold knotted solution to the Euler-Lagrange equations.
This is a member of a family of solutions to the Euler Lagrange
equations that knots in this particular way.  It represents one of the
more strongly ``writhed'' members of this particular class of
solutions.  The ``fins'' on this curve trace out the imbedded $x$ and
$y$ axes, and thus depict the twisting of the rod.}
\label{fig:fiveloop}
\end{figure}

It is possible to generate solutions to the Euler Lagrange equations
for the extremization of the elastic energy that close and that are,
in addition, knotted.  Again, one simply alters the requirement on the
way in which the Euler angle $\phi$ advances over the course of a
``period'' in the oscillation of the quantity $ u(s) = \cos \theta(s)$
between its two limiting values of $b$ and $c$.  Imposing the
requirement that the change in $\phi(s)$ is equal to $m \pi/n$, where
$m$ and $n$ are relatively prime, and $m<n$, one obtains knotted
solutions.  For example, setting $n=3$ and $m=2$, one generates closed
loops in the form of trefoils.  An example of this solution to the
Euler-Lagrange equations is displayed in Figure \ref{fig:trefoil}.
Alternatively, setting $n=5$ and $m=3$, one obtains a knotted solution
with fivefold symmetry.  This solution is shown in Figure
\ref{fig:fiveloop}.  Both displayed loops are members of a family
having the same topological characteristics.  These closed extremal
curves interpolate between nearly circular, ``braided'' loops and
relatively strongly writhing forms, such as are displayed in the
figures.

It is possible to assess the stability of the solutions shown in the
figure. In this case the constraints on fluctuations do not suffice
to stabilize them against deformations that lower their elastic
energy. Thus, in the absence of externally imposed stabilizing
mechanisms, these extremal configurations are not mechanically
stable.

The trefoil solution has also been generated in finite element
calculations \cite{olsen}, and knotted configurations of closed DNA
have long been known to exist {\em in vivo} \cite{knots}

\section{Entropic Corrections}

The analysis of the fluctuations about the classical solution
described abve readily
lends itself to a calculation of the entropic contributions to the
partition function of a loop of DNA at finite temperature. Performing
an expansion of the Euler angles about the form taken in a solution
to the Euler Lagrange equation and retaining terms that are second
order in the deviation of those angles about their classical values,
one obtains the following expression for the partition function of a
closed DNA segment
\begin{equation}
Z \propto e^{-\beta E_{el}\left(\theta_{0}, \phi_{0},
\psi_{0}\right)}\int {\cal D} \gamma(s) \exp\left[- \beta \langle
\gamma(s) |{\cal L}|\gamma(s)\rangle \right]
	\label{partition1}
\end{equation}
where the operator ${\cal L}$ is given by
\begin{equation}
{\cal L} = - \frac{d^{2}}{ds^{2}} +V(s)
	\label{Ldef}
\end{equation}
with $V(s)$ as defined in Eq. (\ref{potential1}). The quantity
$\gamma(s)$ is the deviation of the Euler angle $\theta(s)$ from its
classical value.

\begin{figure}
\centerline{\epsfig{file=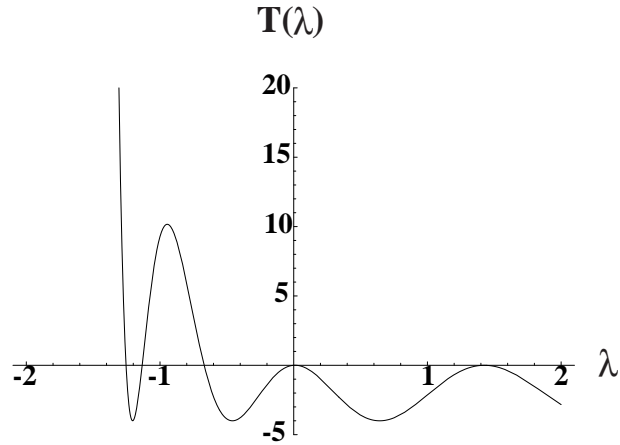,height=3in}}
\caption{The function $T(\lambda)$, as defined by Eq. (\ref{Tdef})
for a characteristic supercoiling solution to the Euler-Lagrange
equations for a closed loop. Note the zeros of this function at
negative values of $\lambda$. The function also passes throught zero
at $\lambda = 0$ as the result of the existence of the translational
mode.}
\label{fig:fred1}
\end{figure}

\begin{figure}
\centerline{\epsfig{file=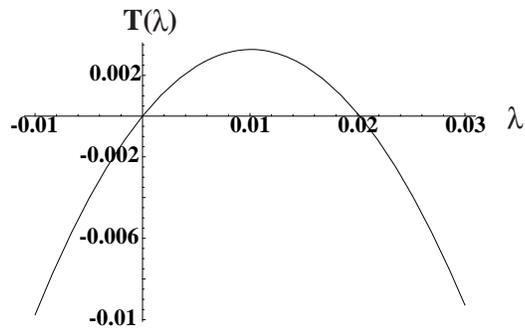,height=2in}}
\caption{A more detailed plot of $T(\lambda)$ in the vicinity of
$\lambda=0$. This function passes through zero with a positive slope,
and there is, in addition, a zero at small, positive, $\lambda$.}
\label{fig:fred2}
\end{figure}

The integral over $\gamma(s)$ in Eq.  (\ref{partition1}) yields the
inverse square root of the Fredholm determinant, $ F(0)$, of the
operator ${\cal L}$, where
\begin{equation}
F(\lambda) = \prod_{l}\left(\lambda_{l}- \lambda \right)
	\label{fredholmdef}
\end{equation}
The quantities $\lambda_{l}$ are the eigenvalues of the operator
${\cal L}$.  The Fredholm determinant is readily calculated with the
use of a method commonly exploited in the study of instanton effects
in nonlinear systems \cite{coleman}.  The application of this method
to the case at hand is outlined in Appendix \ref{sec:fredholm}.  One
finds that the  determinant can be expressed in terms of
the quantity $T(\lambda)$, defined in Eq.  (\ref{Tdef}).  This
quantity is plotted for a characteristic member of family of deformed
loops that interpolate between a circle and a figure eight in Figures
\ref{fig:fred1} and \ref{fig:fred2}.  A noteworthy property of the
Fredholm determinant as displayed in the plots is the fact that this
function of $\lambda$ possesses a zero at $\lambda=0$---this as a
consequence of the existence of the translation mode---and the fact
that there are zeros at negative values of $\lambda$.  These latter
features point to the instability of the unconstrained solution to the
Euler-Lagrange equations.

Given that $T(0)$, and, by extension, $F(0)$, is equal to zero, one
expect the gaussian integration over the variable $\gamma(s)$ to blow
up. However, the translational mode has a well-defined influence on
the fluctuation spectrum. It simply gives rise to a multiplicative
factor reflecting the freedom one has in the fixing location the distortion
on the loop. One eliminates this zero from the Fredholm determinant by
noting that
\begin{equation}
\left. \frac{d}{d \lambda}\prod_{l}\left( \lambda_{l}- \lambda
\right)\right|_{\lambda - \lambda_{l_{0}}} = -\prod_{l \neq
l_{0}}\left(\lambda_{l} - \lambda_{l_{0}}\right)
	\label{deriv}
\end{equation}
This means that we can eliminate the translational mode from the
Fredholm determinant by making the replacement
\begin{equation}
T(0) \rightarrow -\left.\frac{d T(\lambda)}{d
\lambda}\right|_{\lambda = 0}  \equiv -T'(0)
	\label{replacement}
\end{equation}
The result of an unconstrained gaussian integration over fluctuations is,
then,
proportional to
\begin{equation}
\beta^{-N/2}\left(-T'(0)\right)^{-1/2}
	\label{gaussian2}
\end{equation}
where $N$ is equal to the number of modes contributing to the
fluctuation spectrum.  In light of the behavior of $T(\lambda)$, as
displayed in Figures \ref{fig:fred1} and \ref{fig:fred2}, this result
is clearly pathological.

\subsection{Constraints}

Of course, one is not allowed to integrate freely over all
fluctuations. The constraints on the influence of those fluctuations
are the same as applied in the linear stability analysis. The
imposition of constraints leads to an additional multiplicative term
\begin{equation}
\left(\det\left(G\left(\lambda=0\right)\right) \right)^{-1/2}
	\label{addmult}
\end{equation}
For the generaly principle underlying this result, see Appendix
\ref{sec:constint}.  The matrix $G$ is a $5 \times 5$ version of the
matrix introduced in Appendix \ref{sec:genconst}.  The determinant of
this matrix has poles at the negative values of $\lambda$ at which the
function $T(\lambda)$ passes through zero.  Figures \ref{fig:det1},
\ref{fig:det1a} and \ref{fig:det2} display the dependence on $\lambda$
of the two quantities that when multiplied together yield this
determinant.  The plots in these figures are for the same solution to
the Euler-Lagrange equation as generated the $T(\lambda)$ displayed in
Figures \ref{fig:fred1} and \ref{fig:fred2}.  The quantity
\begin{equation}
\beta^{-N/2}\left(-T'(0)\det\left(G\left(\lambda=0\right)\right)
\right)^{-1/2}
	\label{finalfluct}
\end{equation}
represents the contributions of fluctuations about the classical
solution, to within readily calcuated numerical factors.  In the
example for which quantities are displayed, the argument of the square
root is finite and positive.

\section{Conclusion}

We have presented a formalism for obtaining the elastic mimima of a
segment of closed DNA subject to a constraint in the linking number.
The methods exploited here have been utilized to construct the family
of deformations interpolating between the ``relaxed'' circular form
taken by such a segment when the molecule is insufficiently under- or
overwound to induce a supercoiling transition and the figure eight
form that represents the threshold of interwinding. The members of this
family are stable with respect to small fluctuations.  The same
methods also give rise to solutions that represent the evolution of
``higher order'' fluctuations about a circle of the twisted loop.
These solutions to the Euler-Lagrange equations for the extremization
of the elastic energy of a closed loop are unstable with respect to
small fluctuations, in that there exist one or more fluctuation modes
that lead to a lowering of the energy with respect to the classical
solution.  It is also possible to construct knotted solutions to the
Euler-Lagrange equations.  These configurations also represent
saddle-point solutions to the extremization equation.  However, as
previously noted, it is possible to envision stabilizing mechanisms
consistent with the known structure of biological sysems.

In this paper, entire closed loops were generated and studied.  The
same method can, in principle also be utilized to construct a truly
``finite'' finite-element-analysis in which elastic models can be
traced out with the use of non-infinitesimal segments.  Given our
experience in the project described herein, we are confident that the
properties of the segments can be controlled and explicitly displayed.
This ought to give rise to a substantial savings in time and effort in
numerical studies of DNA structure.

\section*{acknowledgements}

The authors would like to acknowledge useful discussions with Zohar
Nussinov and Professor James White. B. Fain acknowledges support from
the A. P. Sloan Foundation.

\begin{appendix}
\section{Stability considerations}
\label{sec:genstab}
\label{app:stability}
Starting with the expression for the elastic energy
\begin{equation}
E_{el} = \int ds {\cal E}_{el}(s)
	\label{el-1}
\end{equation}
where
\begin{equation}
{\cal E}_{el}(s) = \frac{A}{2}\left[ \left(\frac{d
\theta(s)}{ds}\right)^{2} + \sin^{2} \theta(s)
\left(\frac{d\phi(s)}{ds}\right)^{2}\right] + \frac{C}{2} \left[
\frac{d \psi(s)}{ds} + \cos \theta(s) \frac{d \phi(s)}{ds} \right]^{2}
	\label{el0}
\end{equation}
See Eq.  (\ref{c_energy_eqn}).  We expand the Euler angles about their
extremizing values as follows
\begin{eqnarray}
\phi(s) & = & \phi_{cl}(s) + \alpha(s) \label{alphadef} \\
\psi(s) & = & \psi_{cl}(s) + \beta(s)
	\label{betadef}  \\
 \theta(s)& = & \theta_{cl}(s) + \gamma(s)
	\label{gammadef}
\end{eqnarray}
The subscript $cl$ stands for the extremizing, or ``classical''
values of the Euler angles. The deviations from these classical
values, $\alpha(s)$, $\beta(s)$ and $\gamma(s)$, are assumed to be
small. Substituting from Eqs. (\ref{alphadef}) - (\ref{gammadef})
into (\ref{el0}), one finds at quadratic order in the deviations from
the extremizing solutions a local energy equal to
\begin{eqnarray}
\frac{A}{2}\left[ \dot{\gamma}(s)^{2} +
\dot{\phi}_{cl}(s)^{2}\gamma(s)^{2}\left[ \cos^{2}\theta_{cl}(s) -
\sin^{2}\theta_{cl}(s)\right]  \right. \nonumber \\ \left.+ 4 \dot{\alpha}(s)
\gamma(s)
\dot{\phi}_{cl} \sin \theta_{cl}(s) \cos \theta_{cl}(s) +
\dot{\alpha}^{2} \sin^{2}\theta_{cl}(s) \right] \nonumber \\
+ \frac{C}{2}\left[ \left[ \dot{\beta}(s) + \dot{\alpha}(s) \cos
\theta_{cl}(s) - \dot{\phi}_{cl}(s) \gamma(s) \sin \theta_{cl}(s)
\right]^{2} \right. \nonumber \\ \left.
- 2 \left[ \dot{\psi}_{cl}(s) + \dot{\phi}_{cl}(s) \right]
\left[\dot{\alpha}(s) \gamma(s) \sin \theta_{cl}(s) + \dot{\phi}(s)
\gamma^{2}(s) \cos \theta_{cl}(s)/2 \right] \right]
\label{secondorder1}
\end{eqnarray}
When it does not lead to confusion, the, the subscripts will be
dropped from the ``classical,'' or extremum, Euler angles.  Because
the classical Euler angles satisfy an extremum equation, there is no
term linear in the deviations.

A cursory investigation Eq. (\ref{secondorder1}) reveals that the
second-order energy depends on $\alpha(s)$ and $\beta(s)$ only
through their derivatives. Minimizing the energy with respect to
these variables, we are left with the following dependence of the
``fluctuation'' energy on the angular variable $\gamma(s)$
\begin{equation}
\frac{A}{2}\left(\frac{d \gamma(s)}{ds}\right)^{2} -
\frac{\gamma(s)^{2}}{2}\frac{\left(J_{\phi}^{2} +
J_{\psi}^{2}\right)\left( 2 \cos ^{2} \theta(s) + 1 \right) - J_{\phi}
J_{\psi}\cos \theta \left( \cos ^{2}\theta + 5 \right)}{A\sin^{4}
\theta}
	\label{secondorder2}
\end{equation}
Recall that the Euler angle $\theta(s)$ in Eq. (\ref{secondorder2}) is
the solution to the minimzation equation.

The equation (\ref{secondorder2}) for the energy of a fluctuation can
be further reduced if one makes use of the relationship between the
quantities $J_{\phi}$ and $J_{\psi}$ and the roots $a$, $b$, and $c$
of the cubic polynomial in Eq. (\ref{Jdefs}). The new form of the
energy is
\begin{eqnarray}
\lefteqn{\frac{1}{2}\left(\frac{d\gamma(s)}{ds}\right)^{2} }
\nonumber \\  -
&&\frac{1}{4} \frac{\left(2-u(s)\right)
\left(1-u(s)\right)^{2}(a+1)(b+1)(c+1) +
\left(2+u(s)\right)\left(1+u(s)\right)^{2}
(a-1)(1-b)(1-c)}{\left(1-u(s)^{2}\right)^{2}} \gamma(s)^{2}
 \nonumber \\
	\label{secondorder3}
\end{eqnarray}
The quantity $u(s)$ satisfies Eq. (\ref{theta_eqn}). This equation is
equivalent to the expectation value of the energy of a particle in a
one-dimensional potential. This expectation value can be expressed in
terms of the eigensolutions and eigenvalues of the corresponding
Schr\"{o}dinger equation. The stationary version of the
Schr\"{o}dinger equation has the form
\begin{equation}
-\frac{d^{2}\Psi(s)}{ds^{2}} + V(s) \Psi(s) = \lambda \Psi(s)
	\label{schrod1}
\end{equation}
where
\begin{eqnarray}
\lefteqn{V(s)=} \nonumber \\
&&\frac{1}{2} \frac{\left(2-u(s)\right)
\left(1-u(s)\right)^{2}(a+1)(b+1)(c+1) +
\left(2+u(s)\right)\left(1+u(s)\right)^{2}
(a-1)(1-b)(1-c)}{\left(1-u(s)^{2}\right)^{2}}
 \nonumber \\
	\label{potential1}
\end{eqnarray}
The stability of a fluctuation is tied to the sign of the eigenvalues
of the stationary Schr\"{o}dinger equation. If all the allowed values
of $\lambda$ are positive, the solution about which fluctuations occur
is positive. On the other hand if there is one or more negative
$\lambda$, then an instability exists.

\subsection{Constraints}

Fluctuations about the classical solution must obey certain
constraints. In particular, they cannot change the following
properties of the loop: \begin{enumerate}
\item[A.] The loop closes smoothly:
\begin{equation}
\int_{0}^{L}\frac{d \phi(s) }{ds}ds = 2 \pi n
	\label{smooth}
\end{equation}
where $n$ is an integer

\item[B.] The net linking number is fixed:
\begin{equation}
\int_{0}^{L}\frac{d \psi(s) }{ds} ds = \mbox{const.}
	\label{link}
\end{equation}

\item[C.] The loop closes in the $x$ direction:
\begin{equation}
\int_{0}^{L}\sin \theta(s) \cos \phi(s) ds = 0
	\label{xclosure}
\end{equation}

\item[D.] The loop closes in the $y$ direction:
\begin{equation}
\int_{0}^{L} \sin \theta(s) \sin \phi(s) ds = 0
	\label{yclosure}
\end{equation}

\item[E.] The loop closes in the $z$ direction:
\begin{equation}
\int_{0}^{L}\cos \theta(s) ds = 0
	\label{zclosure1}
\end{equation}

\end{enumerate}
In the above expressions, the Euler angles are not necessarily equal
to their extremum values.

Expanding the solution about its extremum form and expressing the
fluctuations in the Euler angles $\phi$ and $\psi$ in terms of the
fluctuation, $\gamma(s)$, in the Euler angle $\theta(s)$,  the
conditions above take the forms shown below \begin{enumerate}

\item[A.] Smooth closure:
\begin{equation}
\sqrt{\frac{F}{2A}}\int_{0}^{L}\frac{p_{1}\left(u(s) - 1 \right)^{2} +
p_{2}\left(u(s) +1 \right)^{2}}{\left(1-u(s) \right)^{3/2}} \gamma(s)
\, ds =0
\label{smooth1}
\end{equation}
Here, and below, the quantity $u(s)$ is $\cos \theta(s)$, where
$\theta(s)$ is the solution to the extremum equations.

\item[B.] Constancy of the link:
\begin{equation}
\sqrt{\frac{F}{2A}}\int_{0}^{L}\frac{p_{1}\left(1-u(s)
\right)^{2}-p_{2}\left(1+u(s)\right)^{2}}{\left(1-u(s)^{2}\right)^{3/2}}
\gamma(s) \, ds =0
	\label{link1}
\end{equation}

\item[C.] Closure in the $x$ direction:
\begin{equation}
\int_{0}^{L}\left\{ u(s) \cos \phi_{cl}(s)
+\sqrt{\frac{F}{2A}}\frac{p_{1}\left(1-u(s)
\right)^{2}+p_{2}\left(1+u(s)
\right)^{2}}{\left(1-u(s)^{2}\right)^{3/2}}{\cal I}_{x}(s) \right\}
\gamma(s) \, ds =0
	\label{xclosure2}
\end{equation}
Here, the quantity ${\cal I}_{x}(s)$ is given by
\begin{equation}
{\cal I}_{x}(s) = \int_{L/2}^{s}\sqrt{1-u(s')^{2}}\sin \phi_{cl}(s')
\, ds'
	\label{Ixdef}
\end{equation}

\item[D.] Closure in the $y$ direction:
\begin{equation}
\int_{0}^{L}\left\{ u(s) \sin \phi_{cl}(s) -\sqrt{\frac{F}{2A}}
\frac{p_{1}\left(1-u(s)
\right)^{2}+p_{2}\left(1+u(s)
\right)^{2}}{\left(1-u(s)^{2}\right)^{3/2}}{\cal I}_{y}(s) \right\}
\gamma(s) \, ds =0
	\label{yclosure2}
\end{equation}
where
\begin{equation}
{\cal I}_{y}(s) = \int_{L/2}^{s}\sqrt{1-u(s')^{2}}\cos \phi_{cl}(s')
\, ds'
	\label{Iydef}
\end{equation}

\item[E.] Closure in the $z$ direction:
\begin{equation}
\int_{0}^{L}\sqrt{1-u(s)^{2}} \gamma(s) \, ds =0
	\label{zclosure2}
\end{equation}

\end{enumerate}

\section{Stability analysis of the circular loop}
\label{sec:circfluc}
The solution of the extremum equations leading to a circular loop is
\begin{eqnarray}
u(s) & = & 0
	\label{circu}  \\
\phi_{cl}(s) & = & \frac{2 \pi s}{L}
	\label{circphi}  \\
\psi_{cl}(s) & = & {\cal L}k\frac{2 \pi s}{L}
	\label{circpsi}
\end{eqnarray}
the quantity ${\cal L}k$ being the linking number of the circular loop.
Making use of Eq. (\ref{secondorder2}), we find for the eigenvalue
equation for fluctuations
\begin{equation}
\frac{A}{2}\left\{-\frac{d^{2}\gamma(s)}{ds^{2}} -\left[\left(\frac{d
\phi_{cl}(s)}{ds}\right)^{2} -
\left(\frac{C}{A}\right)^{2}\left(\frac{d
\psi_{cl}(s)}{ds}\right)^{2} \right] \gamma(s) \right\}= \lambda \gamma(s)
	\label{circfluc1}
\end{equation}
substituting for the rate of change of the classical Euler angles,
$\phi_{cl}(s)$ and $\phi_{cl}(s)$, the eigenvalue equation becomes
\begin{equation}
\frac{A}{2} \left\{ -\frac{d^{2}\gamma(s)}{ds^{2}} - \left(\frac{2
\pi}{L}\right)^{2}\left[ 1+\left(\frac{C}{A}\right)^{2}{\cal L}k^{2} \right]
\right\}
\gamma(s) = \lambda \gamma(s)
	\label{circfluc2}
\end{equation}

Now, the solutions to the equation above are
\begin{equation}
\gamma(s) \propto \cos
\left(\frac{2 \pi n}{L} s + \delta \right)
\label{sinemodes}
\end{equation}
where $n$ is an integer and $\delta$ is an arbitrary phase angle.
Looking at Eq.  (\ref{circfluc2}), one might be tempted to conclude
that the circular loop is always unstable, in that solutions of the
form (\ref{sinemodes}) with $n=0$ and $n=1$ give rise to negative
eigenvalues. However, those solutions are inconsistent with the
constraints on fluctuations. A cursory inspection of the constraints
listed in Eqs. (\ref{smooth1}) to (\ref{zclosure}) reveals that the
following conditions must hold in the case of the circular loop
\begin{eqnarray}
\int_{0}^{L}\gamma(s) \, ds & = & 0
	\label{const1}  \\
\int_{0}^{L} \gamma(s) \cos \left(\frac{2 \pi s}{L}\right) ds & = & 0
	\label{const2}  \\
\int_{0}^{L} \gamma(s) \sin \left(\frac{2 \pi s}{L}\right) ds & = & 0
	\label{const3}
\end{eqnarray}
These three constraints on explicitly rule out flucutations of the
form (\ref{sinemodes}) with $n=0$ or $n=1$. All other values
of $n$ are allowed. If we substitute a fluctuation given by
(\ref{sinemodes}) with $n \ge 2$, the equation for the eigenvalue
$\lambda$ is
\begin{equation}
\lambda = \frac{A}{2}\left(\frac{2 \pi}{L}\right)^{2} \left[ n^{2}-1
-\left(\frac{C}{A}\right)^{2}{\cal L}k^{2} \right]
	\label{nge2}
\end{equation}
The lowest value of $\lambda$ is associated with the smallest allowed value
of $n^{2}$, corresponding to $n=2$. Replacing $n$ by 2 in Eq.
(\ref{nge2}), we find
\begin{equation}
\lambda = \frac{A}{2}\left(\frac{2
\pi}{L}\right)^{2}\left[3-\left(\frac{C}{A}\right)^{2}{\cal
L}k^{2}\right]
	\label{neq2}
\end{equation}
According to Eq. (\ref{neq2}), $\lambda$ will be negative,
corresponding to an instability in the circular configuration, when
${\cal L}k > \sqrt{3}(A/C)$.

\section{General Effect of Constraints on a Stability Calculation}
\label{sec:genconst}

The question of the stability of a solution to the Euler-Lagrange
equations is posed in terms of the eigenvalue spectrum of a linear
operator. This, in turn, can be recast in terms of the problem of
finding extremal values for the expectation value
\begin{equation}
\langle \xi |{\cal L}| \xi \rangle
 	\label{exp1}
\end{equation}
where ${\cal L}$ is the linear operator. The constraints are equivalent to
requiring that the $\xi$ between which the operator is sandwiched is
orthogonal to a set of $m$ $\chi$'s. There is also the constraint on
the absolute magnitude of $\xi$. The constraints are, then of the form
\begin{eqnarray}
\langle \xi|\xi \rangle & = & 1
	\label{normalization1}  \\
\langle \xi | \chi_{l}\rangle & = & 0
	\label{orthogonal1}
\end{eqnarray}
In Eq. (\ref{orthogonal1}), the index $l$ runs from 1 to $m$. The
equation for the extremum of the quadratic form (\ref{exp1}), subject
to the constraints (\ref{normalization1}) and (\ref{orthogonal1}),
takes the form
\begin{equation}
{\cal L} |\xi\rangle = \lambda| \xi \rangle +
\sum_{l=1}^{m}\Lambda_{l}| \chi_{l} \rangle
	\label{extrem1}
\end{equation}
The coefficients $\lambda$ and $\Lambda_{l}$ are Lagrange multipliers,
which enforce the constraints to which the system is subject. The
solution to the above equation is
\begin{equation}
|\xi\rangle = \sum_{l=1}^{m}\frac{\Lambda_{l}}{{\cal L}-
\lambda}|\chi_{l}\rangle
	\label{extrem2}
\end{equation}
The Lagrange multipliers $\Lambda_{l}$ must now be adjusted to ensure
the orthogonality requirements. These requirements are of the form
\begin{eqnarray}
0&=&\sum_{l=1}^{m}\Lambda_{l}\langle
\chi_{k}|\frac{1}{{\cal L}-\lambda}|\chi_{l}\rangle \nonumber \\
&\equiv& G_{kl}\Lambda_{l}
	\label{extrem3}
\end{eqnarray}
This set of $m$ equations for the Lagrange multipliers $\Lambda_{l}$
has non-trivial solutions only if the determinant of the $m\times m$
matrix $G$ is zero. The equation $\left|G_{jk}\right|=0$ represents a
condition on the parameter $\lambda$.

Now, given a solution to Eq. (\ref{extrem3}), we take the expectation
value $\langle \xi {\cal L} \xi \rangle$. Substituting from the right hand
side of Eq. (\ref{extrem3}), we find for this expectation value
\begin{eqnarray}
\sum_{l=1}^{m}\langle \xi |{\cal L} \frac{\Lambda_{l}}{{\cal L}- \lambda}
|\chi_{l}\rangle &=& \sum_{l=1}^{m}\langle \xi |\left( {\cal L}- \lambda
\right) \frac{\Lambda_{l}}{{\cal L}-\lambda}| \chi_{l}\rangle +
\lambda\sum_{l=1}^{m} \langle \xi |\frac{\Lambda_{l}}{{\cal L}-\lambda}
\chi_{l}|\rangle \nonumber \\
&=& \sum_{l=1}^{m}\Lambda_{l}\langle \xi |\chi_{l} \rangle + \lambda
\langle \xi | \xi \rangle \nonumber \\
&=& \lambda
	\label{expect2}
\end{eqnarray}
In Eq. (\ref{expect2}) we have made use of the orthogonality of $\xi$
to the $\chi_{l}$'s. We are also assuming that the function $\xi$ is
normalized.  Thus, in  solving for the value of $\lambda$ that satisfies
Eq. (\ref{extrem3}) we are also determining the effective values of the
eigenvalues of the constrained problem.

\section{The response Kernel}
\label{sec:resp}
The quantity $1/({\cal L}-\lambda)$ represents the response kernel in the
interval $-L/2\le s \le L/2$. This kernel, which can be written in
the form $K(s,s')$, is the inverse of the operator
$-d^{2}/ds^{2} + V(s)$ on that interval, and it has the additional
property that it maps onto periodic functions.  This response is
constructed out of two solutions to the differential equation
\begin{equation}
-\frac{d^{2}\Phi(s)}{ds^{2}} + V(s) \Phi(s) = \lambda \Phi(s)
	\label{diff1}
\end{equation}
The first solution, $\Phi_{1}(s)$ is even under reflection
about the origin. It has the property
\begin{eqnarray}
\Phi_{1}(0) & = & 1
	\label{phi10}  \\
\left.\frac{d \Phi_{1}(s)}{ds}\right|_{s=0} & = & 0
	\label{phi1p0}
\end{eqnarray}
The second solution, $\Phi_{2}(s)$, is odd with respect under
reflection about the origin. It satisfies the following conditions
\begin{eqnarray}
\Phi_{2}(0) & = & 0
	\label{phi20}  \\
\left. \frac{d \Phi_{2}(s)}{ds}\right|_{s=0} & = & 1
	\label{phi2p0}
\end{eqnarray}

That the response kernel  below satisfies all requirements above can
be established by explicit calculation:
\begin{eqnarray}
K(s,s') &=& \Phi_{1}(s_{>})\Phi_{2}(s_{<}) \nonumber \\
&&-\frac{1}{2}\left[ \Phi_{1}(s) \Phi_{2}(s') + \Phi_{2}(s)
\Phi_{1}(s') \right] \nonumber \\
&&+\frac{\dot{\Phi}_{2}(L/2)}{2\dot{\Phi}_{1}(L/2)}\Phi_{1}(s)\Phi_{1}(s')
+ \frac{\Phi_{1}(L/2)}{\Phi_{2}(L/2)}\Phi_{2}(s) \Phi_{2}(s')
	\label{kernel}
\end{eqnarray}
In Eq. (\ref{kernel}), the argument $s_{>(<)}$ is the greater
(lesser) of  $s$, $s'$, and the dots refer to differentiation with
respect to arclength, $s$.

\section{The construction of the Fredholm determinant}
\label{sec:fredholm}
The Fredholm determinant of the operator ${\cal L} - \lambda $, where
\begin{equation}
{\cal L} = -\frac{d^{2}}{ds^{2}} + V(s)
	\label{Ldef1}
\end{equation}
is of the form
\begin{equation}
\prod_{l}\left(\lambda_{l}- \lambda \right)
	\label{freda1}
\end{equation}
where the $\lambda_{l}$'s are the eigenvalues of the operator $L$. In
the case of interest the eigenfunctions of this operator are periodic,
in that if ${\cal L}\Psi_{l} = \lambda_{l}\Psi_{l}$, then $\Psi_{l}(s+L) =
\Psi_{l}(s)$.

The Fredholm determinant is defined in terms of its analytic
properties in the complex $\lambda$ plane.  It is obvious that this
function of the variable $\lambda$ has no poles at finite values of
$\lambda$, and that all zeros are on the real axis, at the locations
of the eigenvalues of the operator ${\cal L}$.  There is another
function of $\lambda$ having this property, formed from the solutions
$\Phi_{1}(s)$ and $\Phi_{2}(s)$, defined in Appendix \ref{sec:resp}.
To construct this function we note that any solution of the equation
${\cal L} \Psi = \lambda \Psi$ can be represented in terms of the two
linearly independent functions $\Phi_{1}$ and $\Phi_{2}$ as follows
\begin{equation}
\Psi(s) = \Psi(0) \Phi_{1}(s) + \left.  \frac{d
\Psi(s)}{ds}\right|_{s=0} \Phi_{2}(s)
	\label{rep1}
\end{equation}
This means that we can express the function $\Psi(s)$ and its
derivative, $\dot{\Psi}(s)$ at $s=L$ in terms of the function and its
derivative at $s=0$ in the following form
\begin{equation}
\left(\begin{array}{l} \Psi(L) \\ \dot{\Psi}(L) \end{array}\right) =
\left(\begin{array}{ll} \Phi_{1}(L) & \Phi_{2}(L) \\
\dot{\Phi}_{1}(L) & \dot{\Phi}_{2}(L) \end{array} \right) \left(
\begin{array}{l} \Psi(0) \\ \dot{\Psi}(0) \end{array} \right)
	\label{transfer1}
\end{equation}
Given the fact that the Wronskian of $\Phi_{1}$ and $\Phi_{2}$ is
equal to one, the matrix on the right hand side of Eq.
(\ref{transfer1}) has a determinant of one.  Now, if the function
$\Psi(s)$ is periodic in $s$, then the right hand side of Eq.
(\ref{transfer1}) is equal to the right hand side.  This leads
immediately to the characteristic equation for the matrix
\begin{eqnarray}
 0 &=& \left|\begin{array}{ll} \Phi_{1}(L) -1 & \Phi_{2}(L) \\
\dot{\Phi}_{1}(L) & \dot{\Phi}_{2}(L) -1 \end{array} \right| \nonumber
\\
&=& 2 - \Phi_{1}(L) - \left.  \frac{d \Phi_{2}(s)}{ds}\right|_{s=L}
\label{characteristic}
\end{eqnarray}
This means that the function of $\lambda$
\begin{equation}
T(\lambda) = -2 + \Phi_{1}(L) + \left.\frac{d \Phi_{2}(s)}{ds}
\right|_{s=L}
\label{Tdef}
\end{equation}
is equal to zero at the eigenvalues of the operator ${\cal L} =
-d^{2}/ds^{2} +V(s)$.  Furthermore, it can be established that this
function is free of singularities and zeros at any other finite value
of $\lambda$.

The function $T(\lambda)$ is not identically equal to the Fredholm
determinant of the operator ${\cal L} - \lambda$, in that the
behavior at $|\lambda| = \infty$ of the two quantities is not the
same. However, if we define $T_{0}(\lambda)$ as the function
corresponding to $T(\lambda)$ when the potential $V(s)$ has been set
equal to zero, and if we denote by $F(\lambda)$ and $F_{0}(\lambda)$
the corresponding Fredholm determinants, then the following
relationship can be esablished
\begin{equation}
\frac{F(\lambda)}{F_{0}(\lambda)}= \frac{T(\lambda)}{T_{0}(\lambda)}
	\label{relationship}
\end{equation}
Given that the Fredholm determinant of the operator ${\cal L}_{0} -
\lambda = -d^{2}/ds^{2} - \lambda$ is readily calculated, Eq.
(\ref{relationship}) leads to a direct determination of the desired
quantity. Figures \ref{fig:fred1} and \ref{fig:fred2} display a
characteristic $T(\lambda)$ for one of the writhing solutions
that interpolate between a circle and a figure eight. Figure
\ref{fig:fred2} shows in detail the behavior of the Fredholm
determinant in the vicinity of $\lambda=0$. Note that this function
passes through zero at $\lambda=0$, and that there is a zero  of this
function at a small, positive value of $\lambda$.

\section{The effect of constraints on a gaussian integral}
\label{sec:constint}

Given the quadratic form
\begin{equation}
\sum_{i,j=1}^{n}x_{i}A_{ij}x_{j} \equiv \vec{x}\cdot {\bf A} \cdot \vec{x}
	\label{quadform}
\end{equation}
the gaussian integral
\begin{equation}
\int
\prod_{i=1}^{n}dx_{i}\exp\left(-\sum_{i,j=1}^{n}x_{i}A_{ij}x_{j}\right)
	\label{gaussint1}
\end{equation}
subject to $m$ the constraints
\begin{equation}
\vec{x}\cdot y_{l}=0 , \ \ \ \  1 \le l \le m
	\label{constraints}
\end{equation}
is equal to
\begin{equation}
\left(\frac{1}{2 \pi}\right)^{m}\int \prod_{l=1}^{m} d
\omega_{l}\left[\int \prod_{i=i}^{n}dx_{i}\exp\left(-\vec{x}\cdot
{\bf A} \cdot \vec{x} + \sum_{l=1}^{m}i \omega_{l} \vec{y}_{l}
\cdot \vec{x} \right) \right]
	\label{gaussint2}
\end{equation}
Performing the integral over $\vec{x}$, we are left with the following
integral over the $\omega_{l}$'s:
\begin{eqnarray}
\left[\det\left({\bf A}/\pi\right) \right]^{-1/2}\left(\frac{1}{2
\pi}\right)^{m}\int \prod_{j=1}^{m}\exp \left( -
\sum_{k,l=1}^{m}\omega_{k}\vec{y}_{k}
\cdot {\bf A}^{-1}\cdot \vec{y}_{l}\omega_{l}/4\right) \nonumber \\
\equiv \left[\det\left({\bf A}/\pi\right) \right]^{-1/2}\left(\frac{1}{2
\pi}\right)^{m}\int \prod_{j=1}^{m}\exp \left( -
\sum_{k,l=1}^{m}\omega_{k} B_{kl}\omega_{l}/4\right)
	\label{gaussint3}
\end{eqnarray}
The gaussian integration over the $\omega_{l}$'s leads to the final
result
\begin{equation}
\left[\det\left({\bf A}/\pi\right) \det \left({\bf B}/4 \pi \right)
\right]^{-1/2}\left(\frac{1}{2 \pi}\right)^{n}
	\label{gaussint4}
\end{equation}

\end{appendix}


\end{document}